\documentclass[aps,prl,twocolumn,showpacs,superscriptaddress,floatfix,citeautoscript]{revtex4-1}

\usepackage{dcolumn}
\usepackage{bm}
\usepackage{amsmath,amssymb}
\usepackage{times}
\usepackage{float}
\usepackage{graphicx, epstopdf}
\usepackage{color}
\usepackage{setspace}

\input alphabet.tex

\begin{document}

\title{High resolution three dimensional structural microscopy by single angle Bragg ptychography}

\author{S. O. Hruszkewycz}
\affiliation{Materials Science Division, Argonne National Laboratory, Argonne, Illinois 60439, USA}
\author{M. Allain}
\affiliation{Aix-Marseille University, CNRS, Centrale Marseille, Institut Fresnel, UMR 7249, 13013 Marseille, France}
\author{M. V. Holt}
\affiliation{Center for Nanoscale Materials, Argonne National Laboratory, Argonne, Illinois 60439, USA}
\author{C. E. Murray}
\affiliation{IBM T.J. Watson Research Center, Yorktown Heights, New York 10598, USA}
\author{J. R. Holt}
\affiliation{IBM Semiconductor Research and Development Center, Hopewell Junction, New York 12533, USA }
\author{P. H. Fuoss}
\affiliation{Materials Science Division, Argonne National Laboratory, Argonne, Illinois 60439, USA}
\author{V. Chamard}
\affiliation{Aix-Marseille University, CNRS, Centrale Marseille, Institut Fresnel, UMR 7249, 13013 Marseille, France}

\date{\today}

\begin{abstract}  
We present an efficient method of imaging 3D nanoscale lattice behavior and strain fields in crystalline materials 
with a new methodology -- three dimensional Bragg projection ptychography (3DBPP).
In this method, the 2D sample structure information encoded in a coherent high-angle Bragg peak measured at a fixed angle is combined with the real-space scanning probe positions to reconstruct the 3D sample structure. 
This work introduces 
 an entirely new means of three dimensional structural imaging of nanoscale materials and eliminates the experimental complexities associated with  rotating nanoscale samples.
We present the framework for the method and demonstrate our approach with a numerical demonstration, an analytical derivation, and an experimental reconstruction of lattice distortions in a component of a nanoelectronic prototype device. 
\end{abstract}

\maketitle 
Inversion methods provide a powerful alternative to traditional objective-lens-based microscopy. Techniques that numerically invert coherent diffraction patterns  into real space images have provided substantial gains in resolution and sensitivity in certain optical, electron, and x-ray microscopy experiments, especially where image-forming lenses are inefficient or difficult to incorporate. The resulting images, formed by inverting reciprocal space diffraction patterns, contain quantitative information that encodes local physical parameters such as permittivity, density, and atomic displacement at sub-beam-size spatial resolutions.

When implemented with hard x-rays, these coherent diffraction imaging (CDI) techniques have enhanced our understanding of the internal structure of nano- and meso-scale materials, especially in operating environments that are difficult to access with other probes. Furthermore, x-ray microscopy methods based on  Bragg diffraction are of particular interest because the sensitivity of x-rays to crystalline distortions in materials can be leveraged to reveal the interplay between structure and properties without disturbing environmental boundary conditions. However, the routine application of inversion methods to coherent hard x-ray Bragg diffraction  is still limited by stringent experimental requirements and long measurement times. 

Given the potential impact of non-destructive 3D structural microscopy and the limitations of current 3D Bragg coherent x-ray inversion methods, advances in  Bragg phase retrieval methods that facilitate the rapid imaging of crystal lattice behavior in realistic environments are critically important.  Here, we introduce a new coherent Bragg diffraction imaging approach, three dimensional Bragg projection ptychography (3DBPP), that provides such a capability. 3DBPP enables 3D image reconstruction from a series of 2D Bragg diffraction patterns measured at a single incident beam  angle, thus forming a new mode of inversion-based 3D strain-sensitive imaging.  As we discuss in this article,  3DBPP is a hybrid real / reciprocal space technique that takes advantage of the high angle of separation between the incident and diffracted beam in a Bragg diffraction geometry and eliminates the need to change the sample angle. In this way, 3DBPP fully exploits the 3D information encoded in a set of 2D Bragg diffraction patterns and paves the way for new high-throughput \emph{in-situ} nanomaterials imaging studies.

\section{Background}

In this work, we demonstrate a method by which fixed-angle 2D coherent diffraction patterns from a focused-beam scanning probe measurement are used to image 3D lattice distortions in nanoscale crystalline structures. 
One key development that we present here is a new reconstruction algorithm that enables this new 3D structural imaging capability.
This  algorithm features concepts utilized in other reconstruction-based microscopy methods, namely ptychography and tomography.
In order to put our 3DBPP method in context, we briefly cover the essential concepts of Bragg coherent diffraction imaging, ptychography, and tomography that relate to our new approach.

X-ray microscopy of nano- and microscale crystals using coherent Bragg diffraction imaging allows three dimensional internal strain fields to be quantitatively measured \cite{nugent10, Nanobook}. In its first implementation, Bragg coherent diffraction imaging involved illuminating a sub-micron-sized isolated crystal with a much larger coherent x-ray beam and measuring the diffracted intensity with an area detector \cite{Pfeifer:2006p69}. The  far-field intensity distribution  in the vicinity of a Bragg peak  was recorded by scanning the fully illuminated crystal through the Bragg \emph{rocking curve} (changing the sample angle relative to the incoming beam), combining successive 2D slices into a 3D Bragg intensity pattern. A three dimensional image of the diffracting crystal can then be numerically retrieved from the measured intensity distribution  using iterative algorithms that constrain the extent of the crystal to within a support that roughly matches its size and shape \cite{Gerchberg72, Sayre52}. This phase retrieval approach has  been successfully used for \emph{in-situ} \cite{Yang:1dc, Watari:2011p5223} and ultrafast \cite{Clark:2013dw}  imaging studies of isolated micro- and nanocrystals. However, the intrinsic limits of the method \cite{Nanobook} constrain its application to a limited class of materials.  For example, extended crystals or samples with large internal strain fields cannot readily be imaged with this approach.

Recently, x-ray Bragg ptychography methods have been developed that eliminate the requirement for isolated crystals and accommodate a broader range of samples. Originally proposed for electron microscopy \cite{Hoppe69, Rodenburg92} and developed extensively with x-rays in the transmission geometry \cite{Rodenburg07, Dierolf10}, the present form of ptychography consists of inverting a set of far-field diffraction intensity patterns collected from overlapping regions of the sample illuminated with a localized beam.  Thus, specific regions of interest can be imaged in continuous samples.  As in the isolated crystal method, 3D images have been reconstructed from a 3D Fourier space intensity distribution about a Bragg peak  \cite{Godard:2011p5158, Berenguer:2013bv}. Although this approach provides access to the internal strain distribution in targeted regions of nano-structured crystals  \cite{Godard:2011}, its implementation requires the measurement of diffraction patterns over a full angular rocking curve at each overlapping scan position, resulting in long measurement times and uncertainties with regard to registration.
 
In general, the structural information encoded in a single far-field coherent diffraction area detector measurement is related to the 3D structure of the sample. In the far-field regime, the components of the diffracted field exhibit a 3D dependence that is related to the illuminated scattering volume  by a Fourier transform.  An x-ray area detector accesses a two-dimensional ``slice'' through the intensity of this 3D reciprocal space distribution.  The slice, defined by the plane of the detector, corresponds to the squared magnitude of the Fourier transform of a set of line integrals through the scattering volume along the direction of the exit beam wave vector.  In other words, the slice is the Fourier transform of a 2D projection of the illuminated sample volume.  This property of x-ray diffraction is known as the ``slice-projection theorem'' (SPT) \cite[Sec. 6.3.3]{Tomobook}, and it enables 2D and 3D imaging by various methods in both  transmission and Bragg reflection geometries \cite{Hruszkewycz:2012wu, Hruszkewycz:2013jp, Takahashi:2013dw}. 

In particular, the slice-projection theorem  forms the foundation of  modern x-ray computed tomography (CT) imaging algorithms.  At its core, CT involves measuring projections of the scattering volume as a function of azimuthal sample angle.  Whether obtained by direct or inversion-based methods, each real-space projection is filtered, then propagated along the  direction of integration, a process known as backprojection. The filtered backprojections at each angle, which represent  each projection in three dimensions in a manner consistent with the SPT, are summed to yield a 3D image \cite{footnoteMA}. However, the quality of a CT reconstruction depends strongly on  angular diversity, and projections must be recorded over nearly the entire  range of sample orientation ($\gtrapprox$ 120$^\circ$) to achieve a robust tomographic  reconstruction \cite{Miao02, Dierolf10}, \cite[Sec. 6.2]{Natterer01b}. 

Below, we introduce a new inversion-based 3D structural imaging method that retrieves the sample without requiring angular diversity. This method relies on the iterative minimization of a cost-function that relates a set of measured 2D high-angle Bragg coherent diffraction patterns to a three dimensional real-space sample structure.

\section{Principles of the 3D Bragg Projection Ptychography}

The key insight of this paper is that, by virtue of the  geometry required to satisfy a crystalline Bragg condition, the angular diversity critical for transmission-geometry CT can be replaced by translational diversity of a localized beam.  We demonstrate that a 3D image can be obtained from only one slice of the rocking curve by presenting numerical and experimental demonstrations as well as an analytical derivation found in the Supplemental.

\begin{figure*}
\includegraphics{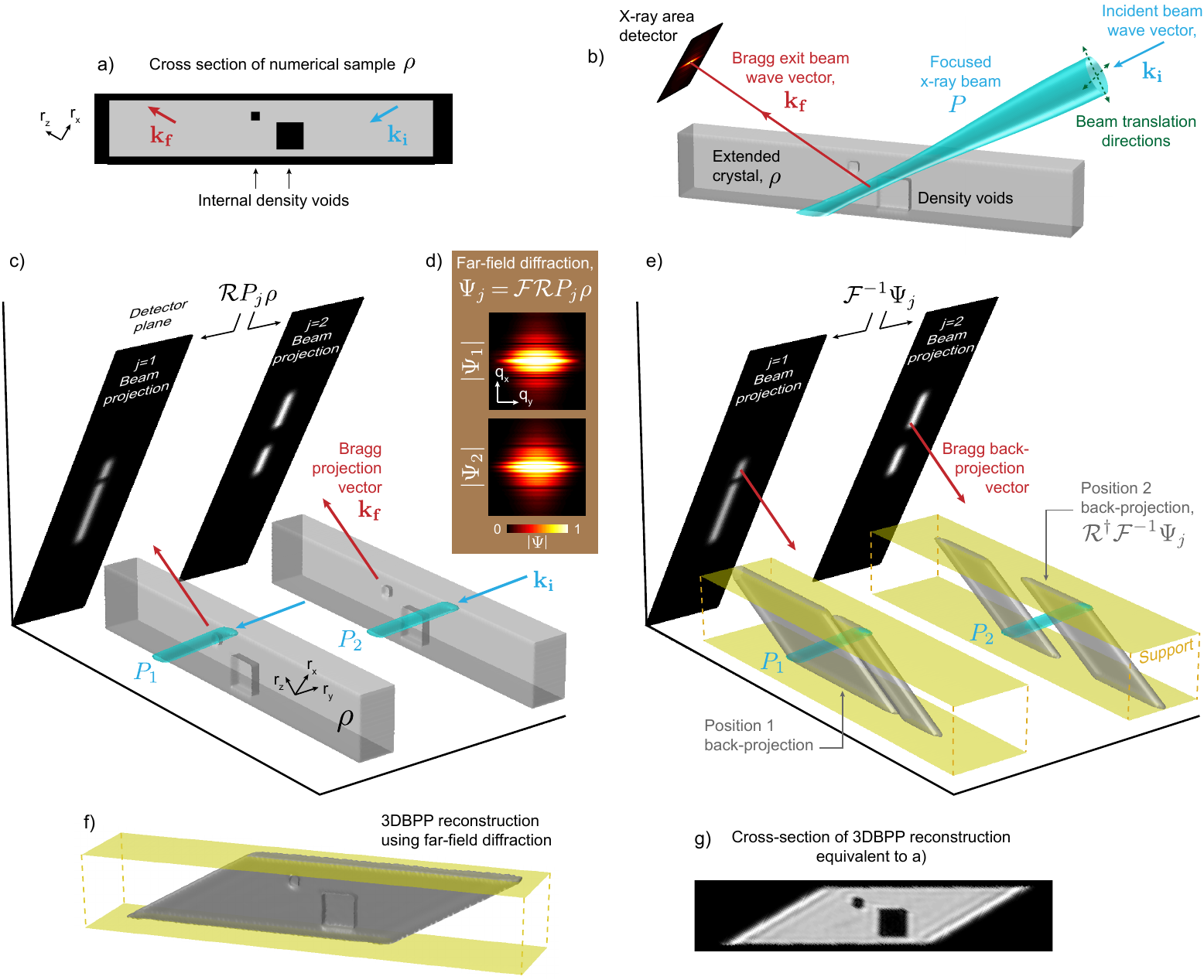}
\caption{
 \textbf{ The principles of 3DBPP.} 
a) To illustrate the different 3DBPP reconstruction operators,  a 3D strain free crystal was generated ($\rho$) containing two internal voids, and a 3DBPP experimental geometry was simulated  (b) using a focussed gaussian beam profile at a high angle Bragg condition defined by the $\bf k_i$ and $\bf k_f$ incident and exit beam vectors. 
We define real space axes $(r_x, r_y, r_z)$ such that ${\bf k_f} \; || \; r_z$ and $(r_x, r_y)$ lie in the detector plane, conjugate to $(q_x, q_y)$ in the far field.
In order to calculate diffraction patterns, projections along ${\bf k_f}$ of the illuminated crystal ($\mathcal{R} P_j \rho$) in the $(r_x, r_y)$ plane are determined (c) at beam positions that intersect  the two voids in $\rho$.  The far-field diffraction amplitudes $| \Psi_j |$ determined by Fourier transforming the projections are shown in d).   The inverse process is shown in e) where, starting from the far-field diffraction $\Psi_j$,  the inverse Fourier transform again yields the 2D  projections of the illuminated crystal. The operator critical to 3DBPP is the back-projection ($\mathcal{R}^\dagger$) of the quantity $\mathcal{F}^{-1} \Psi_j$ along the $\bf k_f$ direction. The backprojection  ``stretches'' the 2D projections along $\bf k_f$ within a 3D support.  Here, the support is made of a pair of planes that limit the extent of the reconstruction in the $z$ direction.  Here, the backprojection intersects  the probe at an oblique angle, effectively localizing the scattering volume.  With 3DBPP,  $\rho$ was reconstructed (f,g) using these operators from a set of 2D coherent Bragg diffraction intensity patterns (See Methods for details). 
}
\label{fig:Radon_principles}
\end{figure*}

At a Bragg condition, the incident and  scattered beam directions are not collinear, but in the hard x-ray regime, they are typically separated by tens of degrees. As a result, the spatial information that is collapsed along the exit beam direction in any individual 2D projection can be encoded by translating a localized x-ray beam in the plane normal to $\bf k_i$, the propagation direction of the incident beam (Figure \ref{fig:Radon_principles}(a, b)).  We use this principle as the basis of our new three dimensional Bragg projection ptychography technique, and we demonstrate that the 3D structure of a diffracting crystal is encoded in, and can be reconstructed from, a set of fixed-angle 2D coherent Bragg peak intensity patterns using spatial diversity. These  patterns are collected by scanning the position of a localized x-ray beam relative to the  crystal such that the beam footprint overlaps between neighboring beam positions. Thus, spatial oversampling is introduced, enabling the use of ptychographic phase retrieval methods while ensuring sensitivity to components of the 3D structure due to the Bragg geometry.  

In 3DBPP, the phases of the diffracted field are retrieved by defining a cost function and gradient that relate the 3D structure of a crystal to the observed 2D Bragg diffraction intensity patterns. At a given beam position, the intensity pattern recorded by the x-ray camera in the far field is the squared amplitude of the diffracted wave field $\Psi_j$:
\begin{equation}
  \langle {I}_j \rangle  =  |\Psi_j |^2,
\label{eq:radon1}
\end{equation}
where ${I}_j$ is the intensity pattern corresponding to the $j$-th beam position.  The intensity pattern recorded in the detector is subject to counting statistics, and so Equation \ref{eq:radon1} defines an expectation value $\langle I_j \rangle$. The diffracted wave $\Psi_j$ is related to the 3D diffracting crystal $\rho$ and the beam $P_j$ (also referred to as the probe) according to \cite{Robinson04,Labat07}:
\begin{equation}
  \Psi_j =  \mathcal{F} \mathcal{R} P_j \rho,
\label{eq:radon2}
\end{equation}
where $\mathcal{F}$ is the (bidimensional) Fourier transform, $\mathcal{R}$ is the (3D) \emph{x-ray projection operator} along the exit beam direction \cite[Sec. II.2]{Natterer01b}.  In this Equation, $P_j$ and $\rho$ are complex-valued three dimensional quantities.  In the Bragg geometry, the phase and amplitude of $\rho$ are related to the Bragg structure factor \cite{Hruszkewycz:2013jp}, which is sensitive to atomic-scale structure in the material. Our aim is to reconstruct $\rho$, thereby imaging the diffraction structure factor in order to quantify  local distortions of the crystal lattice. 

For 3DBPP phase retrieval, an inversion algorithm is used to reconstruct $\rho$ from a set of $J$ intensity patterns $\{I_j\}_{j=1}^J$. The numerical approaches that have successfully been used in transmission geometry ptychography experiments (i.e., difference MAP  \cite{aElser03,aThibault08}, ordered-subset (OS) / Ptychographic Iterative Engine algorithm (PIE) \cite{aRodenburg04,Maiden:2009p4474,Godard:2012uw} or gradient-type iterations \cite{aGuizar08,Thibault12,Godard:2012uw}) can be adapted to accommodate  3DBPP by incorporating a new gradient based on the cost function $\mathcal{Q}(\rho)$:
\begin{equation}
  \mathcal{Q}(\rho) = \sum_{j=1}^J \mathcal{Q}_j(\rho) \quad \text{with} \quad
  \mathcal{Q}_j(\rho) := || \,|\Psi_j (\rho)| - \sqrt{I_j}  \,||^2.
  \label{eq:ErrorMetric}
\end{equation}
In 3DBPP, this cost function yields the following gradient $\partial_j $ for each probe position:
\begin{equation}
\partial_j =  P_j^*  \mathcal{R}^\dagger \mathcal{F}^{-1} \left( \Psi_j - \sqrt I_j \frac{\Psi_j}{|\Psi_j |} \right),
\qquad j=1\cdots J,
\label{eq:grad}
\end{equation}
 where '$*$' is the conjugate operator and $\mathcal{R}^\dagger$ is the adjoint (i.e. \textit{backprojection}) operator \cite[Eq. 2.31]{Natterer01b} associated with the forward-projection operator $\mathcal{R}$. In this work, we incorporated the gradient in Equation \ref{eq:grad} into an iterative OS/PIE algorithm \cite{SCG_blurb} that progressively lowered the cost function $\mathcal{Q}$ and reconstructed an image of the diffracting crystal $\rho$. (See Methods for more details.)

As opposed to previous 3D Bragg ptychography approaches \cite{Godard:2011p5158, Berenguer:2013bv}, this new method is not one of pure Fourier synthesis in which the properties of the 3D reconstruction depend strictly on the measurement and sampling of a 3D volume of reciprocal space. Rather, the 2D structural information of the sample that is encoded in the $q_x$, $q_y$ detector plane is used together with the real-space scanning probe positions to reconstruct the 3D sample structure. Thus, a successful 3DBPP reconstruction must simultaneously retrieve two related spatial components encoded in the  phases of the diffraction patterns. First, each individual intensity pattern, when phased, gives access to the projected scattering volume for each beam position (the quantity $\mathcal{F}^{-1} \Psi_j$ in Figure 1).  Second, the relative phase relationship between the different intensity patterns governs the intersection of each backprojection with the probe at each measurement position. This intersection contains the spatial information along the projection direction that cannot be retrieved from an  individual pattern. To aid in determining the latter phase relationship, various position-referencing constraints may be implemented depending on the geometry of the sample. The thin film reconstructions featured in this work were constrained with a support that confined the extent of the object along the surface-normal direction (as shown in Figure 1). For other samples, different position-referencing constraints may be implemented that serve the same function. 

A numerical demonstration of 3DBPP is featured in Figure \ref{fig:Radon_principles}. We define a 3D thin-film object of uniform density, and introduced two cubic voids in the center of the film. A numerical set of 2D coherent diffraction patterns was generated using Equation \ref{eq:radon2} by scanning a purely real gaussian beam through the central region of the film in overlapping steps. Using 3DBPP implemented with an OS/PIE inversion algorithm, the 3D internal structure of the film in the scanned field of view was successfully reconstructed from this set of 2D patterns (Figure \ref{fig:Radon_principles}(f, g)), thus demonstrating the viability of our approach (see Methods for more details on this numerical test). 

\section{Experimental Demonstration}
\begin{figure*}
\includegraphics{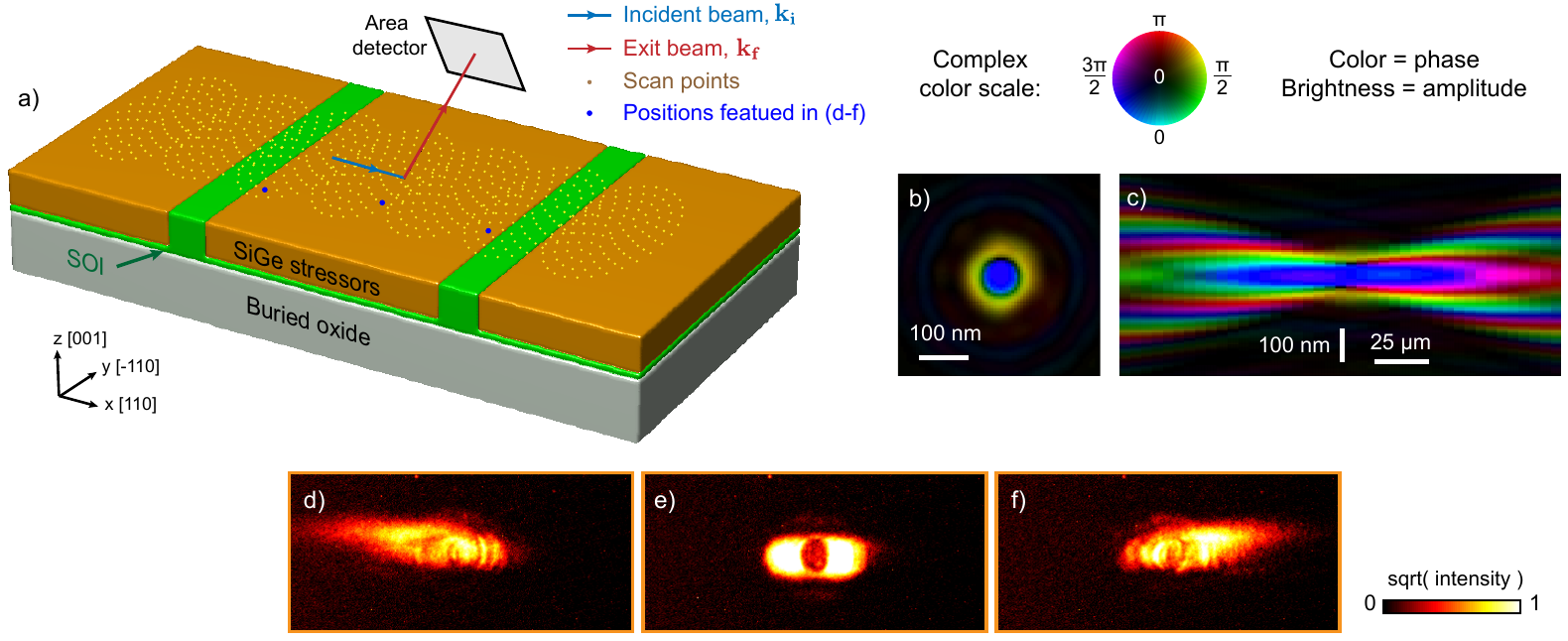}
\caption{\textbf{Experimental geometry.} (a) A schematic of the device architecture is shown featuring the embedded SiGe stressor geometry.  For 3DBPP, coherent diffraction patterns were measured at the specular 004 SiGe Bragg peak at a set of beam positions that effectively overlapped in a series of spiral patterns.  The diffraction patterns from all 707 positions were phased simultaneously with 3DBPP.  (b,c) Prior to diffracting from the eSiGe, the beam wavefront was reconstructed using Fresnel ptychography of a test pattern in the transmission geometry. (d-f) Examples of 004 coherent Bragg intensity patterns used for 3DBPP imaging are shown from different regions of the stressor indicated in (a).  Patterns (d-f) correspond to the positions indicated with blue dots in (a) from left to right.  See Methods and Ref \cite{MartinPRL} for more details.}
\label{fig:setup}
\end{figure*}

To demonstrate the efficacy and potential of 3D Bragg projection ptychography, we imaged the internal lattice displacement field within a  sub-micron-scale crystalline component of a semiconductor prototype device (Figure \ref{fig:setup}). The device structure consisted of periodic embedded SiGe (eSiGe) crystals 460 nm in width that were epitaxially grown to a thickness of 65 nm interstitially between 60-nm-wide linear silicon-on-insulator (SOI) channels (shown in Figure \ref{fig:setup}(a)). This sample design and processing (see Methods) resulted in a complex internal stress state and an accompanying strain field that evolves within the eSiGe stressors in the $x$ and $z$ directions, and is self-similar along $y$ \cite{MartinPRL}.  In particular, the variation of the eSiGe strain field as a function of depth -- variations to which 2D BPP is insensitive \cite{MartinPRL} -- has a direct impact on the electron mobility within the SOI channels in this system and is a critical parameter for nanoscale device engineering.  The 3D reconstruction of an eSiGe strain field presented here via 3DBPP demonstrates the power of this method for in-situ non-destructive structural imaging of functional crystalline nanomaterials.

Coherent nanodiffraction patterns were measured from several adjacent eSiGe stressor crystals  with a zone-plate-focused hard x-ray beam  at a symmetric 004 Bragg condition in which the angle between $\bf k_i$ and $\bf k_f$ was about 60$^\circ$ (see Methods and Figure \ref{fig:setup}).
We note here that the experimental data collection methodology of 3DBPP is exactly the same as that of previously reported 2D BPP experiments \cite{Hruszkewycz:2012wu, Hruszkewycz:2013jp, Takahashi:2013dw, MartinPRL}, however using our new imaging concept, 3D images can now be reconstructed from data that previously yielded 2D projection images.
 Using these Bragg diffraction patterns, a 3DBPP reconstruction was generated using  an iterative algorithm that incorporated the gradient in Equation \ref{eq:grad} into an ordered subset / PIE framework \cite{Godard:2012uw, Maiden:2009p4474}. Because the coherent diffraction observed in the vicinity of the eSiGe 004 Bragg condition (denoted by the reciprocal space vector $\bf G_{004}$) was well separated from scattering from the other components of the device (i.e. SOI and substrate), the  reconstruction represents only the stressor structures.  To aid in the determination of the relative phase relationship of the intensity patterns, a 90-nm-thick support that confined the extent of the reconstruction along $z$ was incorporated into the reconstruction.  This support  constrained the intersection of each backprojection and respective probe function to within a physically realistic pair of parallel planes surrounding the 65-nm-thick film (see Methods).

\begin{figure*}
\includegraphics{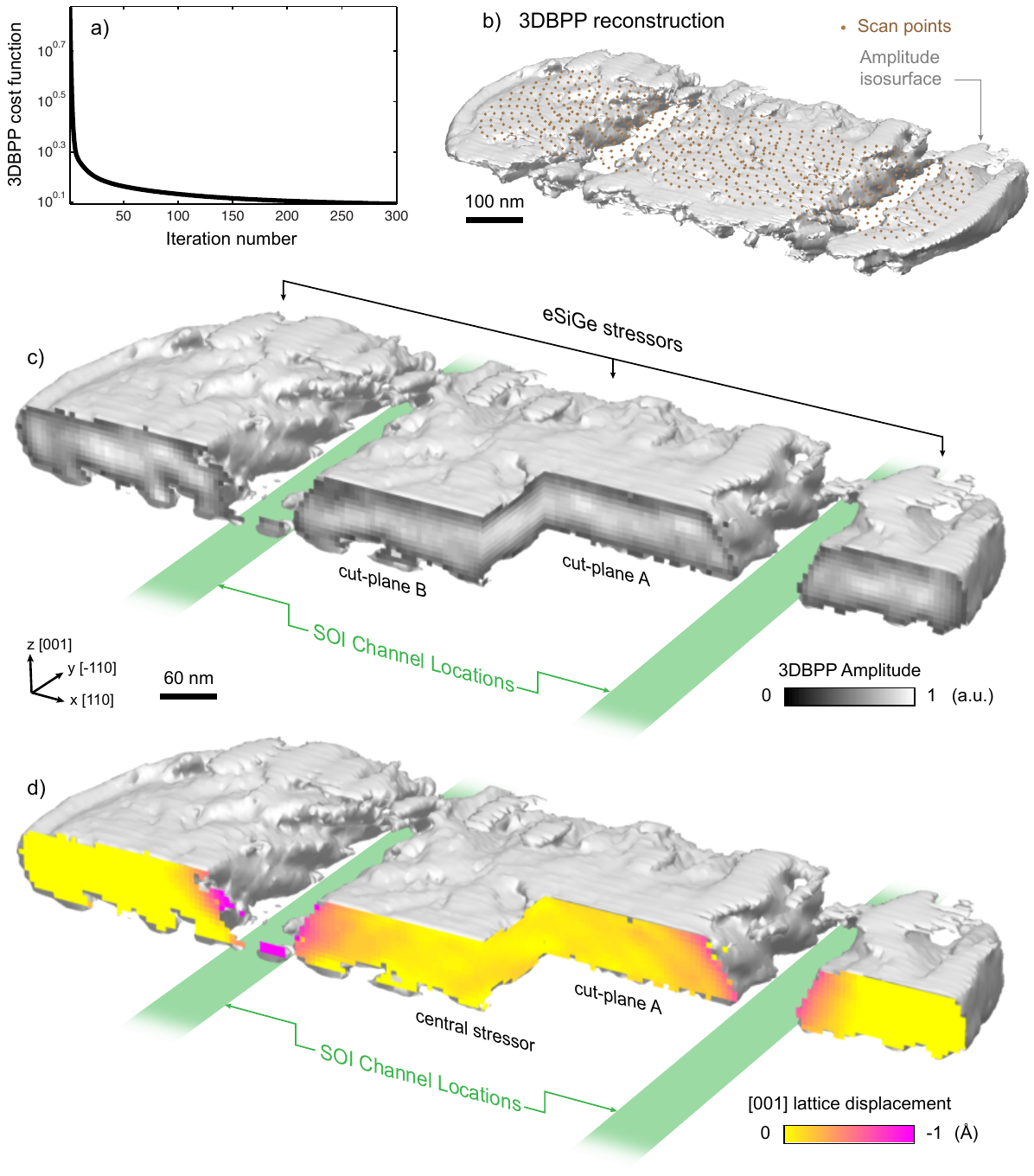}
\caption{\textbf{3D Bragg projection ptychography experimental results} Coherent diffraction patterns measured at the 004 SiGe Bragg peak were phased using 300 iterations of the OS/PIE algorithm adapted for 3DBPP.  (a) The cost function, $\mathcal{Q}(\rho)$, is shown as a function of iteration number.  (b) The isosurface of the amplitude of the resulting reconstruction, showing gaps at the positions of the SOI channels.  The internal structure of the SiGe material in the field of view is revealed by way of cuts through the isosurface depicting the amplitude of the reconstruction, which is closely related to the material density (c), as well as the crystal lattice displacement along the [001] direction (d) which is derived from the phase of the reconstruction.}
\label{fig:sige}
\end{figure*}

The resulting eSiGe reconstruction is shown in Figures \ref{fig:sige}. Two linear SOI channels that are each 60 nm wide run along the $y$ direction. The footprints of these channels are depicted in Figure 3(c,d) by two parallel green strips. The SiGe reconstruction in our 3D field of view therefore features four vertical SiGe/SOI interfaces (along the edges of the green strips) as well as the interface beneath the SiGe stressor material.  The outer boundary of the isosurface shown in Figure 3(a) was defined by the field of view of our 2D beam position scan in the $x$, $y$ plane. 

As is expected for strained crystals, the reconstructed 3D quantity $\rho$ is a complex-valued function, $\rho = |\rho| \exp i \phi$. The amplitude $|\rho|$ is directly related to the electron density of the crystal and the  phase $\phi $ is sensitive to a component of the displacement field in the crystal given by $\phi = \bf G_{004} \cdot \bf u $,  where $\bf u $ is the crystalline displacement field relative to an arbitrary crystalline reference \cite{Pfeifer:2006p69}.

The external shape of the reconstructed amplitude is shown in Figure \ref{fig:sige}b as an iso-surface corresponding to 27\% of the maximum amplitude value.  The reconstructed shape shows the eSiGe features expected in the scanned field of view, including the two missing strips in amplitude corresponding to the positions of SOI channels. The internal structure of the reconstructed stressors is shown in Figure \ref{fig:sige}(c,d).  The vertical cuts through the reconstruction show that the amplitude, which is closely related to the crystal density, is mostly homogeneous with well defined in-plane edges where the eSiGe meets the SOI (further discussion below). However, in this case, the unique strength of Bragg peak inversion lies in visualizing local lattice distortions in a crystal by way of the reconstructed phase.  For  this stressor, the displacement of the crystal lattice along the [001] direction, $\bf{u}_{001}$, was derived from the phase of the 3DBPP reconstruction ($\bf{u_{001}} = \phi / |\bf{G_{004}}|$), and is shown in Figure \ref{fig:sige}(d). As expected for this device architecture,  $\bf u_{001}$ varies in the ($x$, $z$) plane.  The key advantage of 3DBPP is demonstrated by the fact that out-of-plane lattice distortions in this plane that require a 3D reconstruction were resolved,  and that this was accomplished with a fixed-angle scanning probe diffraction measurement.

A crucial issue in this study is the quantification of the resolution obtained in the reconstruction, as this will dictate the applicability of 3DBPP towards the measurement of lattice distortions in different nanoscale crystalline systems. To accomplish this, we used a method similar to that used in Ref \cite{shapiro05}, which is based on an analysis of the complex reconstructed density $\rho$ in terms of spatial frequencies that persist above noise, adapted for a continuous extended 3D sample (see Methods). In the plane of the detector,  anisotropic spatial resolutions of $\sim$18 nm along $x$ and $\sim$17 nm in the  detector direction orthogonal to $x$ were obtained, while along the projection direction (\emph{i.e.} along $\bf k_f$), the resolution was $\sim$25 nm.  The differences in these resolutions serve to highlight the resolution anisotropy that can occur with this technique, though we note that sub-beam-size resolution were achieved in all three directions (see Supplemental).
As in other ptychography experiments, better knowledge of the beam profile and beam positions yields a more accurate and higher resolution 3DBPP image.  Strategies to mitigate uncertainties in these quantities can potentially be adapted from transmission geometry ptychography to 3DBPP in order to improve image fidelity.

\begin{figure*}
\includegraphics{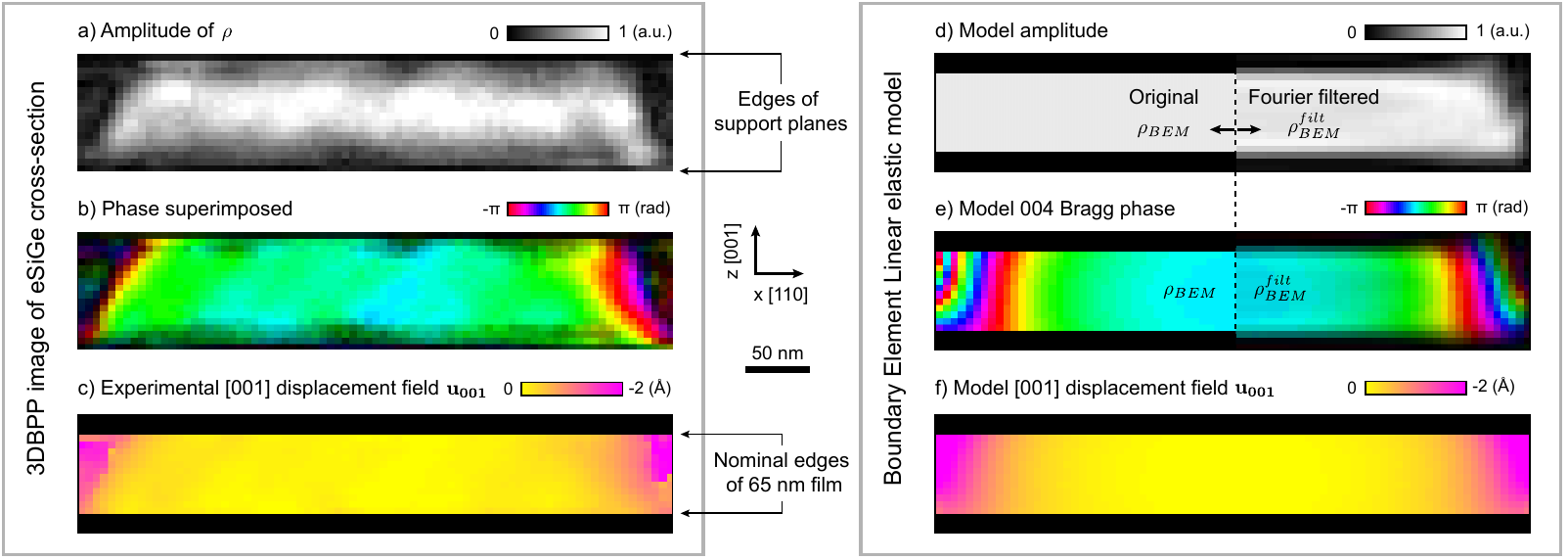}
\caption{\textbf{Comparison with linear elastic model:} A cut through the eSiGe stressor reconstruction is shown as both amplitude (a) and phase (b), with the edges of the 90-nm-thick support planes indicated.  The cross section shown corresponds to cut-plane A in Figure \ref{fig:sige}(c).  In (c), the phases in (b) were converted to units of lattice displacement along the [001] direction ($\bf u_{001}$) confined within an envelope representing the nominal edges of the 65-nm-thick film.  For comparison, a boundary element method (BEM) linear elastic model of $ \bf u_{001}$ in the stressor (f) was converted to the complex density expected for 004 Bragg diffraction, $\rho_{BEM}$.  The amplitude and phase of $\rho_{BEM}$  and $\rho_{BEM}^{filt}$ are shown in (a,b) before and after application of a low-bandpass Fourier filter that replicates the experimental signal level.}
\label{fig:model}
\end{figure*}

The structural fidelity of the method was evaluated by comparing a cross-section of the reconstruction with a linear elastic boundary element method (BEM) model of the eSiGe stressor under  the nominal mechanical boundary conditions of the device  \cite{Murray:2008p5837}. Figure \ref{fig:model} (a-c) shows the  amplitude, phase, and displacement field maps from an ($x,z$) cross-section of the central reconstructed stressor (cut-plane A in Figure \ref{fig:sige}) alongside those of the corresponding BEM model. The BEM model shows that the out-of-plane component of the displacement field $\bf u_{001}$ evolves  within the ($x, z$) plane, intensifying at the top corners to a value of approximately $- 2.5 \; \mathrm{\AA}$ relative to the center  (Figure \ref{fig:model} (f)).  
This variation in $\bf u_{001}$ is the result of an elastic response of
the SiGe lattice due to a change from a near-biaxially stressed state at the center of the stressor to a more uniaxially stressed state at the eSiGe / SOI interface.
The model $\bf u_{001}$ displacement was converted to a complex crystal density $\rho_{BEM}$ calculated for the 004 Bragg diffraction condition, and a Fourier filter was applied to determine phase and amplitude distributions ($\rho_{BEM}^{filt}$). These model distributions are consistent with the experimental Bragg diffraction signal levels (see Methods).  As seen in Figure \ref{fig:model}(d), the Fourier filter modified the rectangular cross-section of the amplitude because  high-spatial-frequency components not experimentally observed in our measurement were filtered out \cite{Godard:2012uw}.  In the center of $\rho_{BEM}^{filt}$, the amplitude remains homogeneous,  but the amplitude envelope falls off near the top corners where the internal displacement field of the crystal is expected to vary more rapidly.  As a result of this displacement field, the phase evolves by $\sim 2 \pi$ in the top corners of the stressor before the amplitude envelope drops off.  These features in $\rho_{BEM}^{filt}$, which come about  from the predicted internal lattice behavior of the stressor as well as the limited dynamical range in the measured intensity, are all observed in the experimental 3DBPP reconstruction (figure \ref{fig:model}a,b). This imaging experiment establishes the effectiveness of our method in using 2D diffraction patterns to reconstruct relevant  3D structural information in targeted regions of nanoscale crystals -- a capability that can be used to explore and discover new phenomena over a broad range of complex strained crystals.

\section{Discussion \& Conclusion }

Our experimental results demonstrate that 3D images of strained crystals can be reconstructed without angular diversity at resolutions appropriate for nanoscale imaging.  However, certain factors must be considered for a successful experiment. Chiefly,  the effectiveness of localizing sample structure along the beam propagation direction (via the $P_j$ and $\mathcal{R}^\dagger$ operators) diminishes as the angle between $\bf k_i$ and $\bf k_f$ approaches zero, while it is maximized at 90$^\circ$.  Secondly, the image resolution along the backprojection direction is  coupled to both the beam size and the step size, while the resolutions parallel to the detector plane are limited by the extent of the reciprocal space signal.  We  note that because the method involves successive backprojections, the reconstruction is susceptible to noise amplification that degrades the image at high iteration numbers in a manner analogous to CT reconstructions with noisy data (see Supplemental). 

Efficient three dimensional microscopy of strain fields in targeted regions of nanoscale crystals under  balanced, undisturbed boundary conditions will enable powerful new studies of \emph{in-situ} materials behavior. By utilizing a new inversion strategy that efficiently uses Bragg diffracted photons from a simpler and dose-efficient experiment, three dimensional Bragg projection ptychography is a significant step in the development of 3D x-ray microscopy, especially with regard to radiation-sensitive materials.  
As compared to Bragg ptychography with rocking curves, 3D images of crystals can be reconstructed using a single diffraction pattern measurement at each point, providing a 50- to 100-fold reduction in dose and acquisition time. Furthermore, our approach simplifies the experimental requirements such that 3DBPP can be implemented on a wide range of coherent synchrotron x-ray beamlines.
More broadly, the integration of projection and backprojection operations into a phase retrieval algorithm can potentially be applied to address challenges in transmission geometry  computed tomography, including the alignment of radial projections with incoherent CT and potentially enabling global 3D inversion of tomographic far-field coherent diffraction patterns.


\section{Methods}

\textbf{Numerical 3DBPP demonstration:} A numerical sample in the shape of a rectangular cuboid (Figure 1) was generated that was $40 \times 30 \times 190$ pixels in dimension with two cubic voids in the sample with respective edge lengths of 7 and 17 pixels.  The numerical object (denoted as $\rho$) was purely real  and had a uniform density outside of the voids.  Figure 1(a) shows a cross-section through the middle of $\rho$, showing the two voids in density.  The  isosurfaces in Figure 1(b,c,f) all depict the density from one outside edge of the object to the central cross-sectional plane shown in (a). Though the entire numerical object is not depicted in the isosurfaces, this visualization helps emphasize and clarify the position and shape of the voids before and after 3DBPP reconstruction.  The object is symmetric about the plane in (a).

The numerical probe used for this example was purely real with a gaussian amplitude profile with a full width at half max width of 4 pixels.  To simulate a high-angle Bragg condition similar to the eSiGe experiment, the incident angle of the probe and the angle of the exit beam ($\bf k_i$ and $\bf k_f$) were both set to 30$^\circ$ with respect to the top surface of the object.  In order to generate a simulated data set, the position of the probe was scanned through the object in a $5\times 61$ point grid with a two pixel step size. At each point $j$, coherent Bragg intensity patterns were calculated according to $\left<I_j\right> = | \mathcal{FR}P_j \rho |^2$.  The resulting noise-free intensity patterns were phased using the OS/PIE 3DBPP algorithm described in the text to generate a reconstruction.  The support used in the reconstruction (yellow surfaces in Figure 1) consisted of parallel planes separated by 40 pixels, matching the dimension of the original object in the $z$ direction.  This support acted only to constrain the top and bottom surfaces of the object,  as  in the experimental eSiGe reconstruction.  The final reconstruction in Figure 1(g) was generated after 200 iterations of the 3DBPP algorithm.  The resolution of this reconstruction along  $\bf k_f$ was found to be 2.6 pixels, demonstrating sub-beam-size resolution of 3DBPP along the projection direction (see Supplemental).

\textbf{Sample preparation:} The sample was a lithographically prepared prototype device \cite{Holt:2013do}, consisting of a series of 460-nm wide trenches  etched into a silicon-on-insulator (SOI) wafer.  The trenches were subsequently filled with epitaxial embedded Si$_{0.8}$Ge$_{0.2}$ (eSiGe) stressors that were 65 nm thick, as confirmed by transmission electron microscopy. The geometry and mechanical boundary conditions of these epitaxial device components are such that strain gradients are expected to build up near the vertical interface between the eSiGe and SOI channel regions.  The device SOI/eSiGE architecture shown in Figure \ref{fig:sige}(a) repeats for $>100$ periods in the $x$ direction, and is invariant and self-similar in the $y$ direction for tens of microns. 

\textbf{Data collection:} The coherent diffraction measurements were done at the Hard X-ray Nanoprobe beamline  \cite{NPblurb, Winarski:2012kv, Hruszkewycz:2014jn}.
A Fresnel zone plate was used to focus 9 keV energy x-rays with wavelength $\lambda = 0.137 $ nm to a $\sim$40 nm diameter focal spot (FWHM intensity of the central focus peak). Prior to collecting nanodiffraction patterns from the eSiGe stressors, the complex wavefront of the beam  was determined in the focal plane with transmission-geometry Fresnel ptychography using a test object \cite{Vine:2009p4586, Thibault:2009p3086}, (Figure \ref{fig:setup}(b-c)).  Using wavefront propagation, the wave field of the beam about the focus was calculated in order to determine the 3D probe function in the Bragg geometry.  

For 3DBPP imaging, coherent nanodiffraction patterns were measured with an area detector at the 004 eSiGe Bragg condition with 10 second exposures. In this symmetric diffraction geometry the incident and exit wave vectors, $\bf k_i$ and $\bf k_f$, were separated by 59.5$^\circ$. The beam was scanned normal to $\bf k_i$ in a series of spiral-like patterns with an effective step size of  $\sim$13 nm separating adjacent points, defining an effective field of view of $\sim 900 \times 200$ nm on the surface of the device.  In total, 707 coherent Bragg diffraction patterns were phased, covering a sample area shown in Figure \ref{fig:setup}(a). Examples of coherent Bragg diffraction patterns from this data set measured at different  positions of a single stressor are shown in Figure \ref{fig:setup}(d-f).  Further details can be found in Ref \cite{MartinPRL}, especially with regard to our implementation of the spiral patterns employed in order to mitigate beam damage. 

\textbf{Inversion procedure:}  The reconstruction was initialized with a 3D starting guess of $\rho$ consisting of random real values. The beam wavefront reconstructed from test pattern data (Figure \ref{fig:setup}(b)) was used to generate a 3D incident focused-beam probe $P$ that was kept constant over the course of the reconstruction.  A total of 300 iterations were performed for the 3DBPP phase retrieval. A support consisting of two parallel planes separated by 90 nm was applied to each new update of $\rho$ such that amplitude outside the volume between the planes was set to zero.  A $\beta$ value of 0.8 was used during the OS/PIE iterations \cite{Maiden:2009p4474}. Similar to the approach in Reference \cite{Godard:2012uw}, we utilized a pre-conditioning matrix to aid in the convergence of the inversion.

\textbf{Estimation of the spatial resolution.}
Following an idea proposed in Reference \cite{shapiro05}, the experimental SiGe reconstruction  ($\widehat{\rho}$) is used to calculate a set of diffraction intensity patterns using the known probe and probe positions. In these diffraction patterns, intensity below a threshold level of $T\sim1$
photon was set to zero resulting in  a set of thresholded intensity patterns $\{I_{j;T}\}_{j=1}^J$.
The spatial resolution of $\widehat{\rho}$ is estimated using $\{I_{j;T}\}_{j=1}^J$ by minimizing the following monodimensional criterion, which is closely related to the criterion expressed in Equation \ref{eq:ErrorMetric}, but which uses the thresholded intensity patterns rather than experimentally observed ones:
\[
r(\alpha) = \sum_{j=1}^J || \,\sqrt{I_{j;T}}(\mathbf{q}) - |\mathcal{F} \mathcal{R} P_j \widehat{\rho}_\alpha(\mathbf{q})| \,||^2.
\]
In this expression, $\widehat{\rho}_\alpha$ is a filtered (low-resolution) version of the reconstruction $\widehat{\rho}$ defined by
its 3D Fourier transform:
\[
\left(\mathcal{F}_{3D} \widehat{\rho}_\alpha \right)(\mathbf{q}) :=
\left\{
  \begin{array}{ll}
   0 &\quad \text{if}~ \left|(\mathcal{F}_{3D} \widehat{\rho}_\alpha)(\mathbf{q})\right| \leq \alpha\\
    \left(\mathcal{F}_{3D} \widehat{\rho} )(\mathbf{q}\right) & \quad \text{otherwise}.
  \end{array}
  \right.
\]
The $\alpha$-filtered version of the reconstruction yields a 3D image that excludes
some of the high spatial frequencies. This criterion is minimized at a value of $\alpha = \alpha_{cr}$, and $\alpha_{cr}$ 
 is then used to determine the spatial resolution of $\widehat{\rho}$. We take the Fourier intensity $| \mathcal{F}_{3D} \widehat{\rho}_{\alpha_{cr}}|$, and determine 
the extent of the non-zero reciprocal space envelope  ($q_{max}$) spanning from the central pixel along each of three directions: the two directions parallel to the plane of the detector and in the direction of projection.  Finally,
$q_{max}$ in each of these directions was converted to an estimate of resolution using the relation: $r=2\pi/(2q_{max})$.

\textbf{Fourier filtered BEM model:} At an exposure time of 10 seconds, the experimental Bragg coherent diffraction signal from the eSiGe stressors fell below the noise level of the detector at  a signal level of $\sim$0.25\% of the maximum observed intensity, below which diffracted signal was not detected. This effectively removed high-frequency components of the diffracted Bragg peak measurement at each position.  Because of the finite measured signal, our experimental reconstruction can be thought of as a low-bandpass filtered version of the reconstruction that would have been obtained with noise-free high-dynamic-range data.

In order to compare the results of the linear elastic model with our experimental reconstruction, a comparable low-bandpass filter was applied. The BEM model, when converted to $\rho_{BEM}$  for the 004 Bragg condition (Figure 4(d,e)), contains high spatial frequency features (sharp edges, rapidly varying phase ramps) that would not all be detected at the signal level corresponding to our experiment.  To represent $\rho_{BEM}$ in a manner that was consistent with the finite experimental signal, frequency components of the Fourier transform of $\rho_{BEM}$ that fell below 0.25\% of the maximum intensity of the FT were zeroed. The inverse Fourier transform of this quantity yielded $\rho_{BEM}^{filt}$, which is an estimate of a 3DBPP reconstruction of the BEM model that  is consistent with the signal-to-noise ratio of our experiment.

\section{Acknowledgements}

This work, including use of the Center for Nanoscale Materials and the Advanced Photon Source was supported by the U. S. Department of Energy, Office of Science, Office of Basic Energy Sciences, under Contract No. DE-AC02-06CH11357.  S.O.H. and P.H.F. were supported by U.S. DOE, Basic Energy Sciences, Materials Sciences and Engineering Division.  V.C. and M.A. were partially funded by the French ANR under project number ANR-11-BS10-0005.  Sample manufacturing was performed by the Research Alliance Teams at various IBM Research and Development facilities. The French OPTITEC cluster is acknowledged for partial support of this work.  The authors also acknowledge Anastasios Pateras for fruitful discussion and help with resolution determination.

\section{Author contributions}
The 3DBPP method was established by SH, MA and VC following the original idea of SH. 
Samples were prepared by CM and JH.
Experimental measurements were performed by SH, MH, CM, and PF.
The manuscript was written by VC, MA and SH with the help of all others.  

\section{Additional information}
The authors declare no competing financial interests.

\newpage 

\noindent{\large \bf Supplemental Section:}

\subsection{Derivation of 3D structural encoding in 3DBPP}

As introduce above, the electronic density of the diffracting crystal and the probe function 
are denoted, respectively, $\rho(\rb)$ and $P(\rb)$. 
For sake of simplicity, we consider the bidimensional case depicted in Figure 1(a). This simplified construction more clearly conveys the fact that structural information in the $r_z$ direction is encoded in a scanning probe Bragg diffraction experiment without a rocking curve.
Our conclusions  remain unchanged when the system is extended to 3D.

The orthonormal reference frame in the real space $(\eb_x,\eb_z)$ is chosen so
that it is the conjugate of the $(\kb_x,\kb_z)$ reference frame, and so that 
 $\kb_x$ is aligned with the detector.  By definition, any $\rb$ (in real space)
and $\qb$ (in reciprocal space) reads
\[
\rb =  r_x\eb_x + r_z\eb_z \quad \text{and} \quad \qb =  q_x\kb_x + q_z\kb_z,
\]
with $||\eb_{x,z}|| = ||\kb_{x,z}||  = 1$.

Let us consider the displacement of the probe
along the direction $\eb_z$ in discrete steps:
\[
P_m(\rb) := P(\rb - m\Delta \eb_z), \quad \text{where} \quad m=0,\cdots M-1.  
\]
Here, the beam step size is defined as $\Delta\in \eR$.
In practice, the displacement of the probe will also have a component along $\eb_x$, but we neglect this component here because it does contribute to the encoding of information along $\eb_z$.

As a result of the Slice Projection Theorem \cite{Natterer01b}[Sec. 6.3.3], when a Bragg condition is satisfied, the coherent far-field intensity pattern observed in the detector is given by:
\begin{equation}
\label{intensity}
I(q_x) = | \tilde{\psi}_m(q_z=0,q_x)  |^2
\end{equation}
where $\tilde{\psi}_m$ is the Fourier transform (indicated by the ``~~$\widetilde{}$~~''  symbol) of the $m$-th sample exit-field: 
\begin{equation}
\label{farfield}
\psi_m(\rb) := P(\rb - m\Delta \eb_z) \rho(\rb) 
\end{equation}
For the purposes of this derivation, we assume that the phases of $\tilde{\psi}_m$ are known (\emph{i.e.} an appropriate phase retrieval strategy was implemented, such as the one we presented in the manuscript).
Thus, we define the following quantity:
\begin{equation}
\label{data}
\tilde{\zeta}(m,q_x) : =  \tilde{\psi}_m(q_z=0,q_x). 
\end{equation}
$\tilde{\zeta}(m,q_x)$ represents a set of complex-valued far-field coherent diffraction wave fields at $M$ different probe positions observed at a fixed sample angle (no rocking curve).
Similarly, in real space $\zeta(m, r_x)$ represents the corresponding set of exit wave functions at the sample at each probe position.
The question we seek to answer is whether structural information about the sample along $\eb_z$ is encoded in either of these quantities. 

To answer this question, we begin by expressing Equation \eqref{farfield} as:
\[
\psi_m(\rb) = \rho(\rb) \times \left[ P(\rb) \otimes \delta(\rb -
  m\Delta \eb_z) \right],
\]
where $\otimes$ is the convolution operator. Taking the Fourier transform leads to:
\begin{widetext}
\begin{eqnarray}
\label{eqar1}
\tilde{\psi}_m(\qb)  &= &\tilde{\rho}(\qb) \otimes \left[ \tilde{P}(\qb)  \times e^{-j2\pi (m\Delta) q_z \langle \eb_z,\kb_z\rangle } \right]\\
\label{eqar2}
& = & e^{-j2\pi (m\Delta) q_z } \int_{\qb'}  \tilde{\rho}(\qb') \tilde{P}(\qb-\qb')  e^{j2\pi (m\Delta)  q_z'} \,\text{d}\qb' 
\end{eqnarray}
\end{widetext}
We now assume that $\rho$ and $P$ are both
\textit{separable} functions, \textit{i.e.,}
\begin{widetext}
\begin{eqnarray}
\rho(\rb) &= & \rho_z(r_z) \times \rho_x(r_x) \quad \Leftrightarrow \quad
\tilde{\rho}(\qb) =  \tilde{\rho}_z(q_z) \times \tilde{\rho}_x(q_x) \\
P(\rb) & = & P_z(r_z) \times P_x(r_x) \quad \Leftrightarrow \quad \tilde{P}(\qb) =  \tilde{P}_z(q_z) \times \tilde{P}_x(q_x) 
\end{eqnarray}
\end{widetext}
This assumption allows us to most clearly demonstrate how spatial information along $\eb_z$ can be encoded without performing a rocking curve. 
To do this, we express Equation \eqref{eqar2} as:
\begin{widetext}
\begin{equation}
\label{eqar2bis}
 \tilde{\psi}_m(\qb)  =  \left( \tilde{\rho}_x \otimes \tilde{P}_x \right) (q_x) \times \left[ \int_{q_z'}
\tilde{\rho}_z(q_z') \tilde{P}_z(q_z-q_z')  e^{j2\pi (m\Delta)
  q_z'} \,\text{d}q_z' \right]e^{-j2\pi (m\Delta) q_z }
\end{equation}
\end{widetext}
According to Equation \eqref{data}, at a single Bragg angle we can only access the far-field diffracted wave field  at $q_z=0$. 
The reciprocal space information we collect as a function of probe position can therefore be expressed as:  
\begin{equation}
\label{databis}
\tilde{\zeta}(m,q_x) = \left( \tilde{\rho}_x \otimes \tilde{P}_x \right)
(q_x) \times h_z(r_z = m\Delta)
\end{equation}
were $h_z$ is the inverse Fourier transform of  $\tilde{h}_z(q) = \tilde{\rho}_z(q) \times \tilde{P}_z(-q)$.
Another convenient expression for Equation \eqref{databis} is:
\begin{equation}
\label{datater}
\tilde{\zeta}(m,q_x) = \left( \tilde{\rho}_x \otimes \tilde{P}_x \right)
(q_x) \times \left( \rho_z \otimes P_z' \right) (r_z = m\Delta)
\end{equation}
 where $P_z' (r_z) := P_z(-r_z)$.  Finally, we can derive the dependence in real space as a set of sample exit waves by taking the inverse Fourier transform along $q_x$:
\begin{equation}
  \label{dataquad}
  \zeta(m,r_x) = P_x(r_x) \times \rho_x(r_x) \times \left( \rho_z \otimes P_z' \right) (r_z = m\Delta).
\end{equation}
From this expression, we see that the sample exit wave field, when measured as a function of probe position $m$, depends on $r_z$ via the
 convolution  of $\rho_z$ with $P_z'$. 
 
 Hence, the structural information contained along the rocking curve direction is encoded in a set of single-angle diffraction 
patterns when they are measured with a scanned focused beam. This conclusion also holds true when a third dimension (into the page) is added. This is because the structural information about the sample along $r_y$ is directly encoded in reciprocal space with an area detector that can access $q_x$ and $q_y$, and does not change the discussion about sample structure along $r_z$.

\subsection{Comparison of Bragg and transmission geometries}

In order to emphasize the role of the Bragg geometry in 3DBPP, we present a  comparison between   images reconstructed from diffraction patterns simulated in both the Bragg and forward scattering geometries (Figure S1). In both cases, intensity patterns give access to a 2D projection of the illuminated volume. However, the relationship between the  exit beam vector $\bf k_f$ (the projection direction) and the incident beam direction $\bf k_i$ in the two cases leads to very different three dimensional reconstructions.  

As outlined in the main text, in the Bragg geometry $\bf k_i$ and $\bf k_f$ are not colinear.  Thus, scanning the localized incident beam in the plane normal to its propagation direction with overlapping steps enables the 3D structure of the sample to be reconstructed using the 3DBPP phasing approach.  In this numerical example, the phases of the far field intensity patterns at each position are known \emph{a-priori}.  Therefore,  a summation of the back-projections at each scan point multiplied by each respective probe can be done: $\sum_j P_{j}^* \mathcal{R}^\dagger \mathcal{F}^{-1} \Psi_j$.  This summation of localized backprojections is shown in Figure S1(b) for the numerical sample described in Figure 1 of the main text.  The resulting image shows that the  three dimensional sample structure can be recovered from a set of 2D projections provided  that $\bf k_i$ and $\bf k_f$  are separated by a non-zero angle typical of hard x-ray Bragg diffraction.  This process yields an image that is a blurred version of the original object, and demonstrates how 3D structural sample information is encoded in a high-angle Bragg ptychography data set without requiring angular diversity.  Starting with the same numerical diffraction information in the form of intensity patterns, 3DBPP was used to recover the phases and reconstruct the object.  As shown in Figure S1(c),  the 3DBPP reconstruction is sharper than the summation of localized back projections, and a sub-beam-size resolution is achieved (see next section in Supplemental) as has been shown in transmission-geometry ptychography experiments \cite{SicairosPRB}.

As compared with the Bragg geometry, $\bf k_i$ and $\bf k_f$ are mostly colinear in a scanning beam transmission geometry diffraction experiment (the maximum angle between $\bf k_i$ and $\bf k_f$ is typically $< 2 ^\circ$), as shown in Figure S1(d).  As a result, no structural information along the propagation direction of the incident beam is encoded in a 2D diffraction pattern measured in this geometry.  This is demonstrated by the fact that both the summation of localized backprojections and the 3DBPP reconstruction of the numerical test object yield images that show the projected density profile of the object stretched along the $\bf k_f$ direction.  In other words, only the backprojection of the original object within the scanned field of view was reconstructed.  In order to recover the three dimensional structure of the sample from such a scattering geometry, scans must be done as a function of sample orientation and computed tomography algorithms must be used for 3D imaging \cite{Dierolf10}.  

\begin{figure*}
\includegraphics{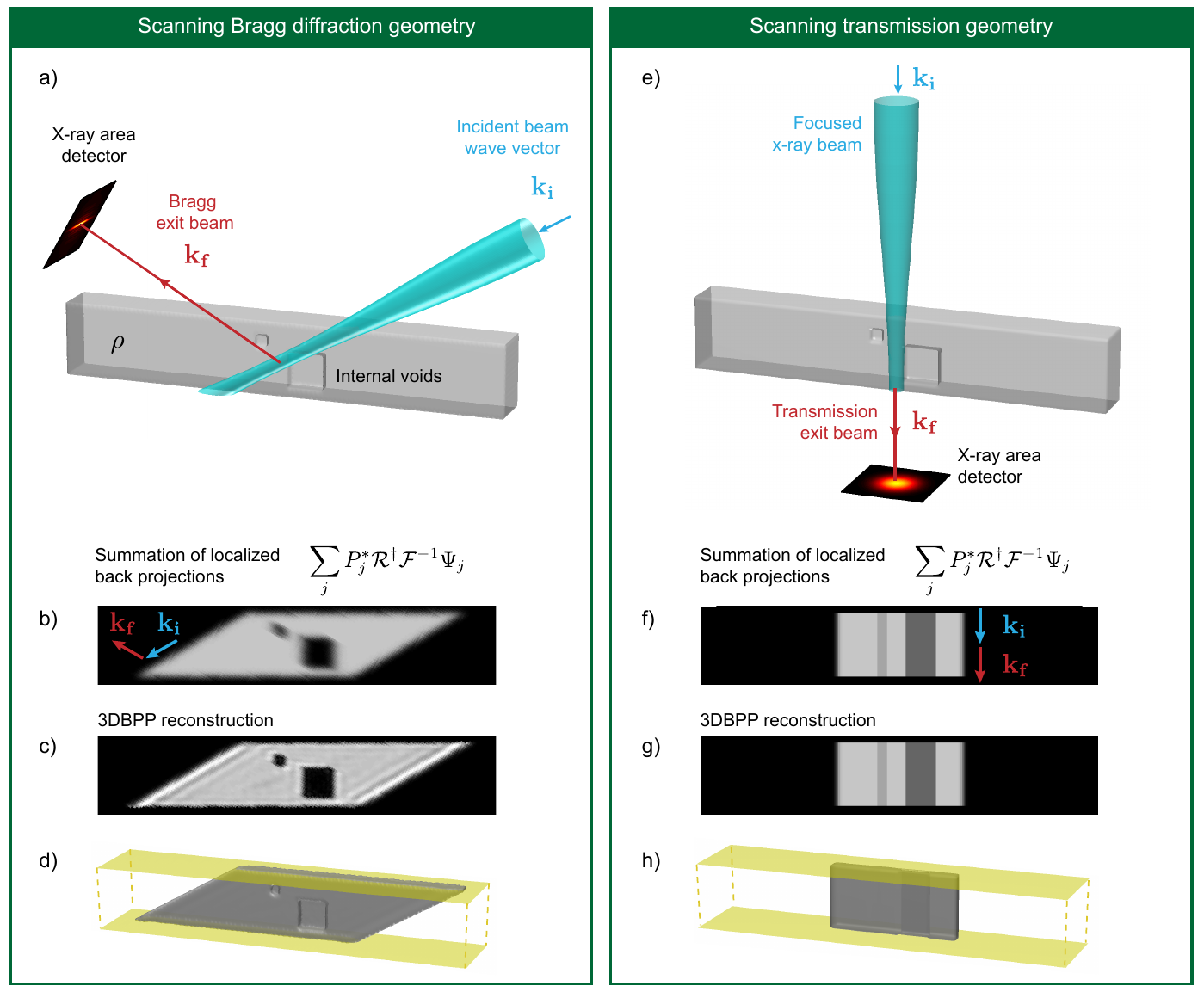}
\caption{ Suppl (S1).
 \textbf{Bragg vs. transmission geometry 3DBPP} 
(a) The Bragg geometry is simulated for 3DBPP of the test object featured in Figure 1 in the main text.  (b) Because the phases of each diffraction pattern are known in this simulation the quantity $\sum_j P_{j}^* \mathcal{R}^\dagger \mathcal{F}^{-1} \Psi_j$ can be calculated, referred to as the summation of localized backprojections.  (c,d) The result of 3DBPP phase retrieval using simulated noise-free diffraction patterns is shown.  e) The same test object is scanned with the same focused beam in a transmission geometry.  (f) The summation of localized backprojections does not accurately represent the voids, nor does the 3DBPP reconstruction (g,h) because  3D sample information is not encoded in a transmission ptychography experiment.
}
\end{figure*}

\subsection{Three-dimensional dependence of resolution}

As in other coherent diffraction imaging techniques, the resolution of a 3DBPP reconstruction varies along different directions. In the plane perpendicular to $\bf k_f$, the resolution is primarily restricted by the extent of the diffraction envelope measured about the Bragg peak, which sets the effective numerical aperture in that plane. This type of $q$-range resolution limitation is typical of diffractive imaging microscopy techniques, and accurately describes the experimental resolution of the eSiGe reconstruction along the axes of the detector plane.




\begin{figure*}
\includegraphics{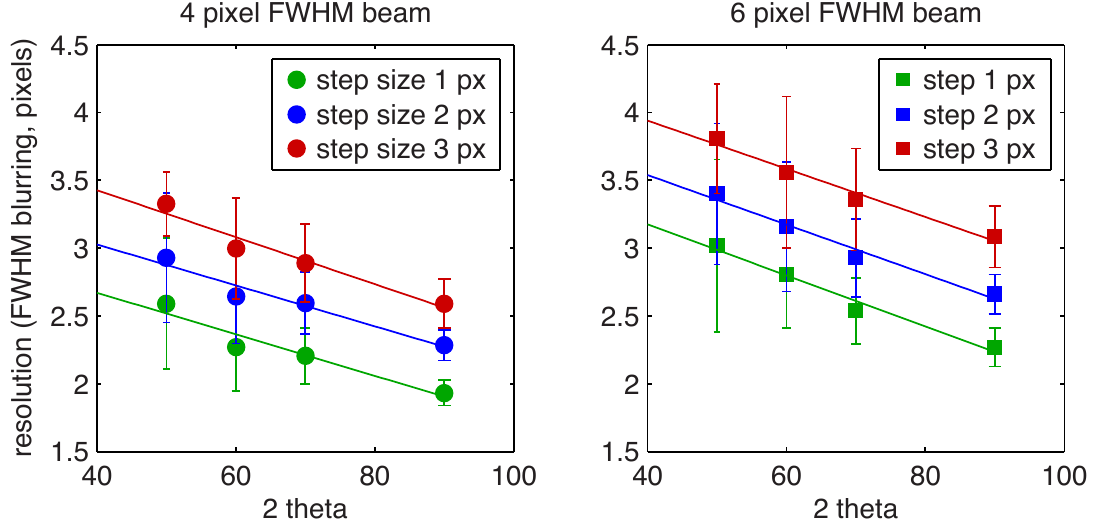}
\caption{Suppl. (S2).
\textbf{Dependence of resolution:}
3DBPP resolution along the $\bf k_f$ direction is shown as a function of step size, scattering angle, and beam size from reconstructions of the numerical object shown in Figure S4(c) at a series of symmetric Bragg reflections.  
Noise-free data was used with a tight support that was the exact outer shape of the original object in order to isolate the effects of the other parameters indicated here. 
The solid lines are linear fits to each data set and serve to highlight the dependence of resolution on Bragg angle, step size, and beam size.
}
\label{fig:res_test}
\end{figure*}

However, the factors that limit image resolution in the projection direction are less obvious.  The resolution in the $\bf k_f$ direction does not stem simply from reciprocal space sampling, but is influenced by the scanning probe nature of the diffraction experiment.  In order to investigate the dependence of resolution along $\bf k_f$ as a function of scattering angle, beam size, and beam step size; 3DBPP reconstructions were done using calculated noise-free data from the  numerical object shown in Figure S4(c).  
Various symmetric Bragg peaks were simulated with $2\theta = \angle {\bf k_i k_f} = 50, 60, 70, 90^\circ$, and an exact support was used that conformed to the outer edges of the sample, leaving the voids to be reconstructed by the algorithm.
Real gaussian beams with a FWHM of 4 and 6 pixels were used in this comparison.
The resolution in the $\bf k_f$ direction of the resulting reconstructions are shown in Figure S2.  They are expressed as the FWHM of a gaussian blur filter kernel applied to a set of amplitude line profiles along $\bf k_f$ through the original object fitted to the corresponding amplitude profiles of the reconstruction.  The mean value of a set of 15 line profile fits from each reconstruction are plotted with error bars of one standard deviation.  Linear fits to the data are shown that are meant to be guides to the eye.  As shown in the Figure, sub-beam-size resolution can be achieved along the $\bf k_f$ direction over an angular range of $2 \theta  \sim 40-90^\circ$ for the conditions surveyed here.  The resolution does, however, have a clear dependence on scattering angle, step size, and beam diameter.  Though a description of the precise dependence and limits of 3DBPP resolution along $\bf k_f$ requires further investigation, based on this numerical study we 
conclude that the resolution of a 3DBPP imaging experiment is highest when the beam diameter and step size are minimized (though these parameters have limits in practice).  Furthermore, the resolution has a strong dependence on scattering angle, which we expect to be optimized when $\bf k_f$ is normal to $\bf k_i$.  As the $2 \theta$ angle approaches 0$^\circ$, we encounter the situation described in Figure S1(e-h), in which all structure along the $\bf k_f$ direction is lost. 

In practice, the dependence and interconnectivity of the 3D resolution function of a 3DBPP reconstruction is undoubtedly a complex function of Bragg angle, beam size, beam amplitude and phase profile, beam step size, experimental positioning accuracy and drift, signal-to-noise ratio, reconstruction algorithm, and other factors.  Furthermore, the experimenter has considerable latitude in defining a support.  In our experimental reconstruction, the support was defined by a pair of parallel planes that were consistent with the expected thickness of the diffracting crystal in the out-of-plane direction. In the case of crystals that have thin-film-like geometries, we find that this is a suitable choice of support.  In addition, the thickness of such crystals can be readily determined by other methods and can even be estimated from the diffraction patterns in the 3DBPP data set.  We note that an effective support for a 3DBPP experiment primarily fulfills the role of a phase reference along $\bf k_f$, and thus can take on forms other than the ones we employed here.

In Figure S3, we explore the dependence of 3D image fidelity as a function of support size by way of numerical simulation of a cubic crystal.  Noise-free intensity patterns were generated from a $7 \times 7$ grid of beam positions illuminating a simple cube with an edge-length of 10 pixels.  The angle between $\bf k_i$ and $\bf k_f$ was 60$^\circ$, as indicated in the Figure.  The plot in Figure S2(a) shows the sum squared difference of pixels between the original cube $\rho_{orig}$ and the resulting reconstruction $\rho$ determined according to $|| \rho- \rho_{orig}|| ^2$ as a function of support size.  The support used in each case was  cubic with edge lengths as shown in the plot.  Two hundred OS/PIE iterations were performed to generate a reconstruction, and reconstructions were done five times in each case.  The mean $|| \rho- \rho_{orig}|| ^2$ value is plotted with error bars of one standard deviation, and values of $|| \rho- \rho_{orig} || ^2$ were determined accounting for the differences in origin of $\rho$ relative to that of the $\rho_{orig}$.  As the edge length of the support increases from being exactly the size of $\rho_{orig}$ to being nearly twice its size, $|| \rho- \rho_{orig}|| ^2$ increases and the reconstructions contain more artifacts that clearly affect the resolution.  However, this test demonstrates that the cube reconstructions are recognizable even with a support that is nearly eight times the volume of the original object, suggesting that the choice of support is rather flexible in 3DBPP.

\begin{figure}
\includegraphics{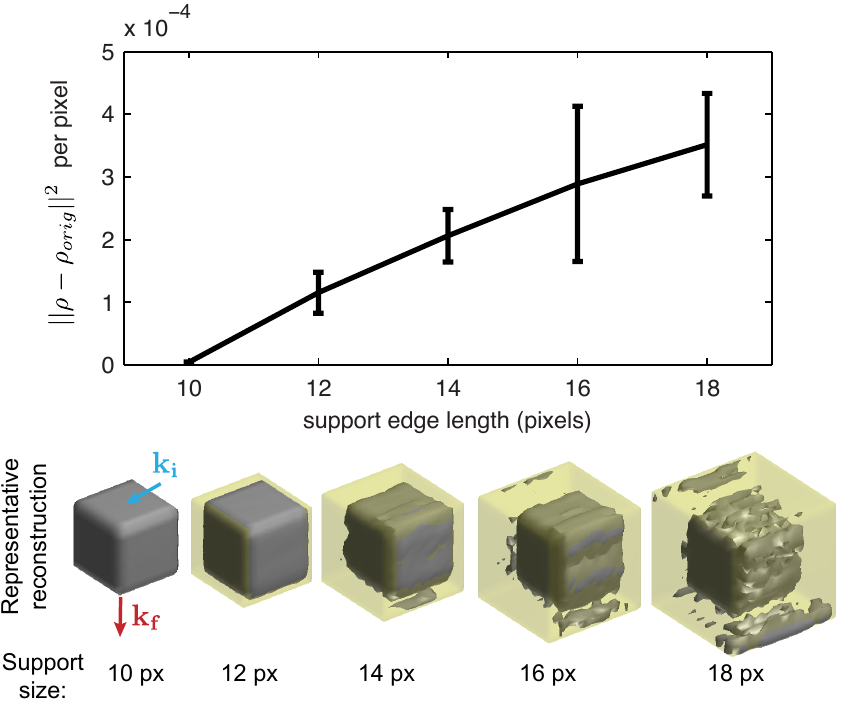}
\caption{ Suppl (S3).
 \textbf{Effect of support in 3DBPP.} 
3DBPP reconstructions of a cubic numerical test object were done using different sized cubic supports.  At each support condition, five separate reconstructions were conducted.  The mean final real-space object error $|| \rho- \rho_{orig}|| ^2$ is shown as  a function of support size in (a).  (b) Examples of a reconstruction from each condition are shown (gray isosurface) along with the corresponding support (yellow surface).  The geometry of the experiment is conveyed by the incident and exit $\bf k_i$ and $\bf k_f$ vectors.
}
\end{figure}

\subsection{Noise amplification during 3DBPP phasing}

In a transmission geometry, the numerical inversion of the x-ray projection operator $\mathcal{R}$ from experimental data, as implemented in CT for example,  is highly sensitive to noise \cite[Sec. 4]{Natterer01b}. This sensitivity is a consequence of the SPT.  In the transmission geometry, the Fourier domain is ``sliced'' in polar coordinates, and high frequency components  are sampled more sparsely as compared to low frequency components when the sample is rotated to measure the 3D intensity distribution.  At a Bragg peak, on the other hand, the 3D intensity distribution can be measured with nearly parallel 2D slices as the sample is rocked in fine angular increments through the crystal rocking curve.  This sampling of 3D reciprocal space at a Bragg peak is therefore done in a cartesian (though not orthogonal) frame \cite{Berenguer:2013bv} rather than a polar coordinate system.  As a result, the high and low frequency components of a 3D Bragg peak intensity distribution are sampled equally, and this leads to a robust ptychographic reconstruction 
when angular diversity is utilized in the Bragg geometry to measure the 3D intensity distribution at each probe position.

 For the 3DBPP technique introduced in this work, however, it is not clear \textit{a priori} if the method 
will be sensitive to noise because the 3D sample information is extracted via shifting
 overlapping probe positions rather than polar or cartesian sampling of a 3D Fourier space far-field intensity distribution. 
Here, we addressed this issue by way of a numerical simulation.  Coherent diffraction data was simulated from a 3D numerical test object (Figure S4) similar to the one featured in Figure 1 in the main text.  While the ptychographic data was simulated in the same way as before,  in this test the resulting intensity patterns were corrupted with Poisson noise in order to study its effects.  From a starting guess of real random numbers, 100 iterations of OS/PIE were done in order to generate a starting estimate of the sample for the purposes of this test.  From this starting estimate, 400 more iterations of the OS/PIE algorithm were performed, and the same object estimate was used as a starting point for 400 conjugate-gradient (CG) iterations.  
The CG inversion has been demonstrated to be convergent  \cite[Sec. 5.2]{Nocedal00}, and hence provides
(in the asymptotic limit) a local minimizer of the cost-function $\mathcal{Q}$ given by Equation 4 in the main text. 
In both cases, the reciprocal space error metric $\mathcal{Q}$ was tracked as a function of iteration number as well as the real space object error metric $|| \rho- \rho_{orig}|| ^2$.

\begin{figure*}
\includegraphics{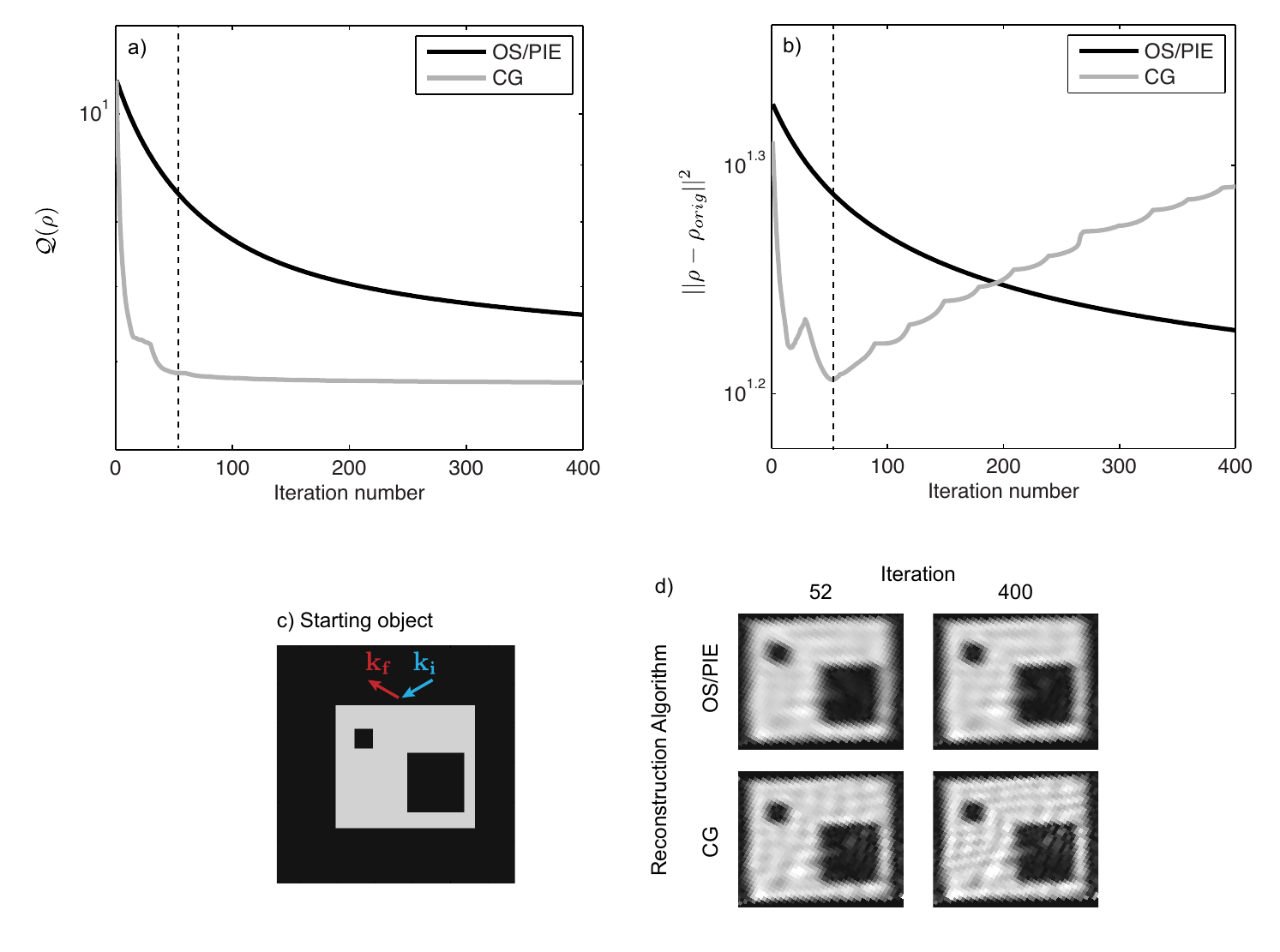}
\caption{Suppl (S4).
 \textbf{3DBPP  in the presence of noise.} 
(a) Evolution of the cost function $\mathcal{Q}$ and of the error metric in the object space $|| \rho- \rho_{orig}|| ^2$ (b)
shown as a function of iteration number  for both the  CG and the OS/PIE. 
(c) The cross-section through the center of the starting object $\rho_{orig}$ and scattering geometry are shown.
(d) Cross sections of the reconstruction $\rho$ are shown at different iterations the CG and OS/PIE.
}
\end{figure*}

As shown in Figures S3(a, b), both algorithms minimize $\mathcal{Q}$, however their behavior is very different when evaluating how closely the reconstruction resembles the original object at each iteration in terms of $|| \rho- \rho_{orig}|| ^2$.  At first, the CG algorithm improves the quality of the result more rapidly than the OS/PIE algorithm (Figure S3(b)), however, after 52 iterations the CG $|| \rho- \rho_{orig}|| ^2$ begins to increase, while $\mathcal{Q}$ continues to decrease.  By contrast, with OS/PIE, the values of $|| \rho- \rho_{orig}|| ^2$ and $\mathcal{Q}$ both decrease over the course of the 400 iterations (however, this is not necessarily  an indication of convergence).  
According to standard inversion theory \cite[Sec. 6]{Bertero98}, the behavior displayed by the CG algorithm suggests  that 3DBPP is  noise sensitive, meaning that high-resolution  features in the object cannot be retrieved from the noisy data set without regularization. However, stable band-limited reconstructions are obtained in the early iterations by the CG, and similar behavior can be obtained by other gradient-based iterative algorithms, including the OS/PIE as shown here. Despite the fact that the OS/PIE approach cannot reach (but will approach) the minimum of $\mathcal{Q}$ using noisy data because of the inherent properties of the algorithm, an OS/PIE-based reconstruction strategy was adopted in the work presented in this paper because it provides  band-limited reconstructions \cite{Godard:2011,Godard:2012uw}.


\begin{thebibliography}{10}
\bibitem{footnoteMA}
Tomography with coherent x-rays has also been demonstrated with far field coherent x-ray Fourier synthesis methods that do not rely on backprojections. Rather, a 3D diffraction patterned assembled from individual 2D slices is phased and inverted \cite{Miao02,Chapman96}.  Also, in the transmission geometry, the 3D nature of the Ewald sphere has been used to extract 3D sample information \cite{MiaoNature}.
\bibitem{NPblurb}
The hard x-ray nanoprobe beam line is operated by the Center for Nanoscale Materials at the Advanced Photon Source.
\end{thebibliography}
\end{document}